\documentclass[12pt]{article}
\usepackage{a4wide,axodraw}
\catcode`@=11
\newcount\@tempcntc
\def\@citex[#1]#2{\if@filesw\immediate\write\@auxout{\string\citation{#2}}\fi
  \@tempcnta\z@\@tempcntb\m@ne\def\@citea{}\@cite{\@for\@citeb:=#2\do
    {\@ifundefined
       {b@\@citeb}{\@citeo\@tempcntb\m@ne\@citea\def\@citea{,}{\bf ?}\@warning
       {Citation `\@citeb' on page \thepage \space undefined}}%
    {\setbox\z@\hbox{\global\@tempcntc0\csname b@\@citeb\endcsname\relax}%
     \ifnum\@tempcntc=\z@ \@citeo\@tempcntb\m@ne
       \@citea\def\@citea{,}\hbox{\csname b@\@citeb\endcsname}%
     \else
      \advance\@tempcntb\@ne
      \ifnum\@tempcntb=\@tempcntc
      \else\advance\@tempcntb\m@ne\@citeo
      \@tempcnta\@tempcntc\@tempcntb\@tempcntc\fi\fi}}\@citeo}{#1}}
\def\@citeo{\ifnum\@tempcnta>\@tempcntb\else\@citea\def\@citea{,}%
  \ifnum\@tempcnta=\@tempcntb\the\@tempcnta\else
   {\advance\@tempcnta\@ne\ifnum\@tempcnta=\@tempcntb \else \def\@citea{--}\fi
    \advance\@tempcnta\m@ne\the\@tempcnta\@citea\the\@tempcntb}\fi\fi}
\catcode`@=12

\newcommand{\bea}{\begin{eqnarray}}
\newcommand{\eea}{\end{eqnarray}}
\newcommand{\be}{\begin{equation}}
\newcommand{\ee}{\end{equation}}
\begin{document}

%
%
%%%%%%%%%%%%%%%%%%%%%%%%%%%%%%%%%%%%%%%%%%%%%%%%%%%%%%%%%%%%%%%%%%%%%%%%%%%%
%%%%%%%%%%%%%%%%%%%%%%%%%%%%%%%%%%%%%%%%%%%%%%%%%%%%%%%%%%%%%%%%%%%%%%%%%%%%
\title{
\vskip-3cm{\baselineskip14pt
\centerline{\normalsize DESY~12--024 \hfill ISSN~0418--9833}
\centerline{\normalsize February 2012\hfill}}
\vskip1.5cm
Counting master integrals: integration-by-parts procedure with effective
mass
}

\author{
{\sc Bernd~A.~Kniehl, Anatoly~V.~Kotikov}\thanks{On leave of absence from
Bogolubov Laboratory of Theoretical Physics,
Joint Institute for Nuclear Research, 141980 Dubna (Moscow Region), Russia.}
\\
\\
{\normalsize II. Institut f\"ur Theoretische Physik, Universit\"at Hamburg,}\\
{\normalsize Luruper Chaussee 149, 22761 Hamburg, Germany}
}

\date{}

\maketitle
%%%%%%%%%%%%%%%%%%%%%%%%%%%%%%%%%%%%%%%%%%%%%%%%%%%%%%%%%%%%%%%%%%%%%%%%%%%%%%%
\abstract{
We show that the new relation between master integrals recently obtained in
Ref.~\cite{KaKn} can be reproduced using the integration-by-parts technique
implemented with an effective mass.
In fact, this relation is recovered as a special case of a whole family of new
relations between master integrals.
\medskip

}

\newpage

%\tableofcontents

%=====================================================================
\renewcommand{\thefootnote}{\arabic{footnote}}
\setcounter{footnote}{0}
%%%%%%%%%%%%%%%%%%%%%%%%%%%%%%%%%%%%%%%%%%%%%%%%%%%%%%%%%%%%%%%%%%%%%%%%%
%\section{Introduction}
%\setcounter{equation}{0}
%%%%%%%%%%%%%%%%%%%%%%%%%%%%%%%%%%%%%%%%%%%%%%%%%%%%%%%%%%%%%%%%%%%%%%%%%%%%

Recently, one of us in collaboration with Mikhail Kalmykov 
found \cite{KaKn} a new relation between some specific Feynman integrals,
which is actually absent in modern computer programs based on 
the integration-by-parts (IBP) technique \cite{ibp} (for a recent review, see
Ref.~\cite{Grozin}). The new relation arises in the framework of the
so-called differential reduction (see Refs.~\cite{KaKn,Bytev:2011ks} and 
references cited therein) developed by these authors during last several years.
This decreases the number of master integrals and, thus, leads to a 
simplification of calculations.

In this short note, we recover this relation directly in the framework of
the IBP technique by introducing an effective mass originating from the
reduction of one-loop integrals to simple propagators (see
Refs.~\cite{Kotikov:1991hm,Fleischer:1997bw} and Eq.~(\ref{I.1}) below).
In fact, this relation is found to be a special case of a whole family of new
relations between master integrals.

%
%  Fig. Sunset with MMm
%
%\vspace*{2.0cm}
%\SetScale{0.8}
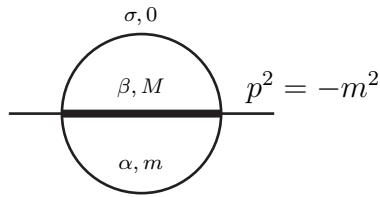
\begin {figure} [htbp]
\centerline{
\begin{picture}(150,100)(0,0)
\SetWidth{1.0}
%\SetWidth{2.0}
\CArc(75,50)(30,0,180)
\CArc(75,50)(30,180,360)
\Line(45,50)(25,50)
\Line(105,50)(125,50)
\SetWidth{3.0}
\Line(45,50)(105,50)
\Text(75,30)[c]{$\scriptstyle \alpha,\,m$}
\Text(75,60)[c]{$\scriptstyle \beta,\,M$}
\Text(75,87)[c]{$\scriptstyle \sigma,\,0$}
\Text(115,60)[l]{$p^2=-m^2$}
%\Text(75,5)[c]{(a)}
\end{picture}
}
\caption{Two-loop sunset diagram $J_{012}(\sigma,\beta,\alpha)$ involving
propagators with masses 0, $M$, and $m$ raised to the powers $\sigma$, $\beta$,
and $\alpha$, respectively, taken on the mass shell $p^2=-m^2$.}
\label{sunsetMMm}
\end{figure}

Following Ref.~\cite{KaKn}, let us consider the two-loop self-energy 
sunset-type diagram $J_{012}$ with on-shell kinematics, defined as 
\begin{equation}
J_{012}(\sigma,\beta,\alpha) = \frac{1}{\pi^{n}}
\int 
\left.
\frac{d^nk_1d^nk_2}{[(p-k_1)^2]^\sigma[(k_1-k_2)^2+M^2]^\beta[k_2^2+m^2]^\alpha}
\right|_{p^2=-m^2},
\label{J012}
\end{equation}
where $n=4-2\varepsilon$ is the dimensionality of space time.
It is depicted in Fig.~\ref{sunsetMMm}. 

Considering the standard Feynman representation of the following one-loop
diagram as a one-fold integral
\begin{eqnarray}
I(\alpha_1,\alpha_2)
&\equiv &  
\frac{1}{\pi^{n/2}}
\int 
\frac{d^n k}{[k^2+M_1^2]^{\alpha_1} [(p-k)^2+M_2^2]^{\alpha_2}}
\nonumber \\
&=& 
\frac{\Gamma(\alpha_1+\alpha_2-n/2)}{\Gamma(\alpha_1)\Gamma(\alpha_2)}
\int^1_0 \frac{ds \, s^{n/2-1-\alpha_1}\overline{s}^{n/2-1-\alpha_2}
}{[p^2 + 
M_1^2/s+M_2^2/\overline{s}]^{\alpha_1+\alpha_2-n/2}},
\label{I.1}
\end{eqnarray}
with $\overline{s}=1-s$,
we can interpret this as an integral over a new propagator with the effective
mass $M_1^2/s+M_2^2/\overline{s}$.
In previous papers \cite{Kotikov:1991hm,Fleischer:1997bw}, this procedure was
used to decrease the numbers of loops in analyses of different types of master 
integrals and, thus, to simplify calculations.

Using Eq.~(\ref{I.1}) with $M_1=M$ and $M_2=0$,
we can represent the considered two-loop diagram
 $J_{012}(\sigma,\beta,\alpha)$ as the one-fold integral
\begin{equation}
J_{012}(\sigma,\beta,\alpha) = 
\frac{\Gamma(\sigma+\beta-n/2)}{
\Gamma(\sigma)\Gamma(\beta)}
\int^1_0 \frac{ds}{s^{\beta+1-n/2}\overline{s}^{\sigma+1-n/2}
} \,
I_{12}(\alpha,\beta+\sigma-n/2) ,
\label{I.2}
\end{equation}
where 
\begin{equation}
I_{12}(\alpha_1,\alpha_2) = \frac{1}{\pi^{n/2}}
\int 
\left. 
\frac{d^n k}{[(p-k)^2+m^2]^{\alpha_1} [k^2+M^2/s]^{\alpha_2}}
\right|_{p^2=-m^2}
\label{I.3}
\end{equation}
is a one-loop on-shell diagram.

Applying the IBP procedure to the one-loop integral 
$I(\alpha_1,\alpha_2)$ considered in Eq.~(\ref{I.1}),
with the distinguished line carrying the index $\alpha_1$  
(see, for example, Ref.~\cite{Kotikov:1991hm}),\footnote{%
I.e.\ the factor coming in IBP procedure has the form 
$n=d(k-k_1)^{\mu}/dk^{\mu}$, where $k-k_1$ is the momentum of the propagator 
carrying the index $\alpha_1$.}
we have the general relation
\begin{eqnarray}
(n-2\alpha_1-\alpha_2) I(\alpha_1,\alpha_2)&=&\alpha_2 \left[
I(\alpha_1-1,\alpha_2+1) 
- \left(p^2+M_1^2+M_2^2\right) 
I(\alpha_1,\alpha_2+1)\right]
\nonumber\\
&&{}- 2\alpha_1 M_1^2 I(\alpha_1+1,\alpha_2) .
\label{I.6a} 
\end{eqnarray}

Thus, for $I_{12}(1,\alpha_2)$ considered in Eq.~(\ref{I.3}), 
we can apply Eq.~(\ref{I.6a}) with $M_1=m$, $M_2=M/\sqrt{s}$, $p^2=-m^2$,
and $\alpha_1=1$.
%, and $\alpha_2=\varepsilon$. 
The result is
\begin{equation}
(n-2-\alpha_2) I_{12}(1,\alpha_2) =
\alpha_2 I_{12}(0,1+\alpha_2) 
- \alpha_2 \frac{M^2}{s} 
I_{12}(1,1+\alpha_2) - 2m^2  I_{12}(2,\alpha_2)  ,
\label{I.6} 
\end{equation}
where the tadpole $I_{12}(0,1+\alpha_2)$ has the form
\begin{equation}
I_{12}(0,1+\alpha_2) = 
\frac{\Gamma(\alpha_2+\varepsilon-1)}{\Gamma(\alpha_2+1)} 
{\left(\frac{s}{M^2}\right)}^{\alpha_2+\varepsilon-1} .
\label{I.7}
\end{equation}

Integrating Eq.~(\ref{I.6}) with $\alpha_2=\sigma+\beta-n/2$, multiplied by
the factor
$$
%\be
\frac{\Gamma(\sigma+\beta-n/2)}{
\Gamma(\sigma)\Gamma(\beta)}\,
\frac{1}{s^{\beta+1-n/2}\overline{s}^{\sigma+1-n/2}}
%\label{I.7}
%\nonumber
$$
%\ee
%$\Gamma(\varepsilon)/[s\overline{s}]^{\varepsilon}$ 
as on the r.h.s.\ of Eq.~(\ref{I.2}), over $s$, 
we obtain
\begin{eqnarray}
(4-3\varepsilon-\sigma-\beta)  J_{012}(\sigma,\beta,1) &=&
%\Gamma(\varepsilon)\Gamma(1-\varepsilon)\Gamma(2\varepsilon-1) 
%(M^2)^{1-2\varepsilon}
%\Gamma(2-\varepsilon-\sigma)\Gamma(\sigma-1+\varepsilon)
\frac{\pi}{\sin[\pi(2-\varepsilon-\sigma)]}\,
\frac{\Gamma(\sigma+\beta-3+2\varepsilon)}{
\Gamma(\sigma)\Gamma(\beta)} 
(M^2)^{3-2\varepsilon-\sigma-\beta}
\nonumber \\
&&{} - M^2 \beta J_{012}(\sigma,1+\beta,1) - 2m^2  
J_{012}(\sigma,\beta,2) .
\label{I.8} 
\end{eqnarray}
Putting $\sigma=\beta=1$, we recover the new relation discovered in
Ref.~\cite{KaKn} as a special case of a more general class of relations
between IBP master integrals.

In conclusion, applying the IBP procedure to a one-loop integral
with an effective mass in one of its propagators, we produced the new 
relation (\ref{I.8}) between ordinary IBP master integrals.
This relation coincides for $\sigma=\beta=1$ with the one recently discovered
in Ref.~\cite{KaKn}, but it is more general and obtained in a more
straightforward way.
We intend to extend this analysis to the case of the off-shell sunset diagrams 
in a future work.
The effective-mass procedure applied here to reduce the number of master
integrals with respect to the one achieved by the ordinary IBP procedure may
in principle be applied whenever the considered topology contains a bubble
subdiagram.
We expect that relations between ordinary IBP master integrals thus obtained
may be usefully implemented in modern computer packages based on the IBP
procedure.

%%%%%%%%%%%%%%%%%%%%%%%%%%%%%%%%%%%%%%%%%%%%%%%%%%%%%%%%%%%%%%%%%%
%%%%%%%%%%%%%%%%%%%%%%%%%%%%%%%%%%%%%%%%%%%%%%%%%%%%%%%%%%%%%%%%%%
\vspace{5mm}
\noindent 
{\bf Acknowledgments}\\
We are grateful to M.~Yu.~Kalmykov for useful discussions.
The work of A.V.K. was supported in part by RFBR Grant No. 10-02-01259-a
and the Heisenberg-Landau program.
This work was supported in part by the German Federal Ministry for Education
and Research BMBF through Grant No.\ 05~HT6GUA, by the German Research
Foundation DFG through the Collaborative Research Centre No.~676
{\it Particles, Strings and the Early Universe---The structure of Matter and
Space Time}, and by the Helmholtz Association HGF through the Helmholtz
Alliance Ha~101 {\it Physics at the Terascale}.
%%%%%%%%%%%%%%%%%%%%%%%%%%%%%%%%%%%%%%%%%%%%%%%%%%%%%%%%%%%%%%%%%%
%%%%%%%%%%%%%%%%%%%%%%%%%%%%%%%%%%%%%%%%%%%%%%%%%%%%%%%%%%%%%%%%%%

\end{document}